\documentclass[12pt,twoside,reqno]{amsart}

\usepackage{amssymb}
\usepackage{amsmath}

\numberwithin{equation}{section}

\def\diam{\text{\rm diam}}

\def\CC{{\mathcal{C}}}

\def\R{{\mathbb R}}
\def\C{{\mathbb{C}}}

\def\e{\varepsilon}

\newtheorem{theorem}{Theorem}[section]
\newtheorem{lemma}[theorem]{Lemma}

\title[Reflectionless measures with a point mass and singular continuous component]
{Reflectionless measures with singular components}

\author{}
\address{}
\email{}
\author{F.~Nazarov, A.~Volberg, P. Yuditskii}
\thanks{ }

\begin{document}
\maketitle

\section{Introduction and main results}

The Cauchy transform (potential) $\CC_{\nu}(z)$ of a Radon measure $\nu$ in $\C$ is defined by
$$
\CC_{\nu}(z):=\int\frac{d\nu(\zeta)}{\zeta-z}\,.
$$

\vspace{.1in}

We consider here only {\bf real} $\nu$ supported on compact $E\subset \R$, and call this measure reflectionless if there exists a limit $\CC_{\nu}(x+i0)$ and it is purely imaginary for $\nu$ almost every $x\in E$. We call this measure weakly reflectionless if if there exists a limit $\CC_{\nu}(x+i0)$ and it is purely imaginary for $dx$ almost every $x\in E$.

We need two classes of compact sets. First is the homogeneous sets:
$|E\cap (x-h,x+h)| \geq \eta h$ for all $x\in E$ and all $h\in (0, \diam E]$.  Such sets have many interesting properties. Among those are:
\begin{itemize}
\item If $\omega$ denotes the harmonic measure of $\Omega:=\C\setminus E$ then $\omega$ is mutually absolutely continuous with
$dx|E$ and $d\omega/dx \in L^p,\, p>1$. The exponent $p$ depends on $\eta$ above.

\item If $\ell$ is any bounded complementary interval of compact $E$ (there is also one unbounded interval ``centered" at infinity), and $c_{\ell}$ is a maximum point for Green's function $G(z) := G(z, \infty)$ of $\Omega$ then 
\begin{equation}
\label{widom}
\sum_{\ell} G(c_{\ell}) <\infty\,.
\end{equation}

\item If $\nu = wdx +\nu_s$ is a real measure, $\nu_s$ its singular part, and  $\CC_{\nu}(x+i0)$ exists and it is purely imaginary for $dx|E$ almost every $x\in E$ ( so $\nu$ is weakly reflectionless) then $\nu_s =0$.

\end{itemize}

These facts can be found in \cite{SoYu}, \cite{Z}, directly or under a small disguise.

The sets which satisfy the property \eqref{widom} alone are called Widom sets (and $\Omega$ is called Widom domain).

Our goal is to show that there exists a Widom set $E$ and a positive measure $\nu = w dx + \delta_{b_0}$ such that $\nu$ is weakly reflectionless. The obvious candidate for weakly reflectionless measure is the harmonic measure.
In fact, here is a theorem which is proved (practically) in \cite{MPV}:
\begin{theorem}
\label{harm1}
Let $E, \omega$ be a compact set on $\R$ and harmonic measure
of $\C\setminus E$ with pole at infinity. We do not assume homogenuity or any other property on $E$. Then $\CC_{\omega}(x+i0)$ exists and it is purely imaginary for $dx|E$ almost every $x\in E$.
So harmonic  measure $\omega$ is {\bf always} weakly reflectionless.
\end{theorem}

Notice that theorem is not empty only if $|E|>0$.

But for Widom domains we cannot hope to have harmonic measure as 
having both a) weak reflectionless property and b) non-trivial singular part. This is because of the

\begin{theorem}
\label{harm2}
For Widom $E$, harmonic measure is always mutually absolutely continuous with respect to $dx|E$.
\end{theorem}

(Of course harmonic measure of anything cannot have a point mass, anyway.)

\section{Example.}
\label{example}

Example. We start with segment $[b_0, a_0]$ on the real axis, and we throw away open intervals $l_j:=(a_j, b_j)$ such that $l_{m}$ is to the left from $l_{m-1}$ and they accumulate to $b_0$ from the right.
We call $l_j$ gaps, and we call $[b_{j+1}, a_j]$ slits. $E:=  \cup_j [b_{j+1}, a_j]\cup\{b_0\}$.  In the future $j$-th slit will be very small with respect to $j$-th gap. Here is a function analytic in $\C\setminus E$:
$$
R(z) = -\frac{1}{\sqrt{(z-a_0)(z-b_0)}}\Pi_{j=1}^{\infty}\frac{z-x_j}{\sqrt{(z-a_j)(z-b_j)}}\,.
$$
Properties of $R$: 1) $R$ is purely imaginary on slits $[b_{j+1}, a_j]$;
2) $R$ is real on gaps $(a_j, b_j)$, and it is positive on $(a_j, x_j)$ and negative on $(x_j, b_j)$; 3) $\Im R(z) \geq 0$ if $\Im z\geq 0$ (Nevanlinna function).

Therefore, 
$$
R(z) =\int\frac{d\mu(x)}{x-z}\,,
$$
 where $\mu$ is a positive measure on $E$.
 
 Notice the following
 
 \begin{equation}
 \label{positive}
 \Im [(z-b_0) R(z)] \geq 0\,\,\,\text{if}\,\,\, \Im z \geq 0\,.
 \end{equation}

In fact,
$$
(z-b_0)\int_E\frac{d\mu(x)}{x-z} = \int_E\frac{(z-x)d\mu(x)}{x-z} + \int_E\frac{(x-b_0)d\mu(x)}{x-z} = -\|\mu\| + \int_E\frac{d\nu(x)}{x-z}\,,
$$
where $\nu= (x-b_0) \mu$ is a positive  measure as $x-b_0\geq 0$ on $E$. This proves \eqref{positive}.

But then $(z-b_0) R(z)$ also has the following property: 

\begin{equation}
\label{positive1}
(x-b_0)R(x) <0,\,\,\, \text{for}\,\,\, x< b_0, \,\,\,\text{or} \,\,\, x>a_0\,.
\end{equation}

Therefore, $\log [-(x-b_0)R(x)]$ has compactly supported imaginary part called $-g(x)$. In fact, 
by \eqref{positive1} $g(x) =0$ on $(-\infty, b_0)\cup (a_0,\infty)$. Then in the upper half-plane $\C_+$ we have 
$$
\log [-(z-b_0)R(z)] = -\frac1{\pi}\int\frac{g(x)\,dx}{x-z}\,.
$$
and hence,

\begin{equation}
\label{m}
(z-b_0)R(z) = -e^{-\frac1{\pi}\int\frac{g(x)\,dx}{x-z}}\,.
\end{equation}

Notice that all these minuses are really essential. By the definition of $g$ we know not only that it has a compact support but also we know that $g=\pi/2$, where $R$ is purely imaginary and that $g=\pi$, when $R$ is negative. Otherwise $g$ is zero. So,
$g \neq 0$ on $E\cup \cup_{j=1}^{\infty}(x_j,b_j)= \cup_{j=1}^{\infty}(x_{j+1},a_j) $.

Now $\mu$ has a point mass at $b_0$ iff 
$$
\lim_{x\rightarrow b_0-} (x-b_0)R(x) >0\,.
$$
We use \eqref{m} to conclude that this may happen iff 
$$
\int\frac{g(x)\,dx}{x-z} <+\infty\,.
$$
 
 But we just calculated $g$, and clearly the last condition is equivalent to
 
 \begin{equation}
 \label{series}
 \sum_j\frac{a_j-x_{j+1}}{x_{j+1}-b_0} <\infty\,.
 \end{equation}
 
We can easily take $x_j= b_j$ for all $j$, and transform that to

\begin{equation}
 \label{series1}
 \sum_j\frac{a_j-b_{j+1}}{b_{j+1}-b_0} <\infty\,.
 \end{equation}

Now we will show that condition \eqref{series1} is compatible with the Widom property of
$\Omega= \C\setminus E= \C\setminus (\cup_{j=1}^{\infty}[b_{j+1},a_j]\cup\{b_0\})$.

Given a domain $\Omega$, its harmonic measure $\omega(E,z)$ means harmonic measure of $E\subset \partial\Omega$ at pole $z\in \Omega$. 

Choose 
$$
b_0=0, a_0=1\,.
$$

We will choose $a_j,b_j, j\geq 1$ inductively. Let $\ell_1=(a_1, b_1),...,\ell_{n-1}=(a_{n-1}, b_{n-1})$ be chosen. Consider $\Omega_{n-1}: = \C \setminus ([b_1,a_0]\cup[b_2,a_1]\cup...\cup
[b_{n-1}, a_{n-2}])$.  We consider also $\Omega_n= \Omega_{n-1} \setminus [b_n, a_{n-1}]$, where $b_n$ is chosen as follows:
we denote corresponding harmonic measures by $\omega_{n-1}, \omega_n$ and require that
\begin{equation}
\label{12}
\omega_n([b_n, a_{n-1}], 0) \geq \frac12\,.
\end{equation}

Now we choose $a_n>0$ and very close to $0$. Then we choose $b_{n+1}$ by the criterion \eqref{12}, namely, to have
\begin{equation}
\label{12again}
\omega_{n+1}([b_{n+1}, a_{n}], 0) \geq \frac12\,.
\end{equation}

Obviously we need

\begin{lemma}
\label{harmonic}
Let $E$ be  a compact subset of the positive half-axis. Let $I=[b',a]$ be a segment between $0$ and $E$, such that for $\Omega= \C\setminus (E\cup I)$
we have $\omega (I, 0)\geq 12$. Then
$$
\lim_{a\rightarrow 0} \frac{b'}{a} =1\,.
$$
\end{lemma}

Suppose the lemma is proved. Then we can finish our construction. In fact, Choose $a_n$ from \eqref{12again} so close to $0$ that for $b_{n+1}$ from \eqref{12again} one has
$$
\frac{b_{n+1}}{a_n} \in (1-2^{-n},1)\,.
$$
Then condition \eqref{series1}: $\sum_j\frac{a_j-b_{j+1}}{b_{j+1}-b_0} <\infty$ is automatically satisfied (recall that $b_0=0$ in our construction).

We are left to check that \eqref{12} for all $n$ ensures that
the domain $\Omega=\lim \Omega_n= \C\setminus ([b_1,a_0]\cup[b_2,a_1]\cup...\cup
[b_{n-1}, a_{n-2}]\cup....)$ is a Widom domain. 

Let $G_n, G$ are Green's function of $\Omega_n, \Omega$ with pole at infinity. Let $c_k$ be the critical points of $G$, $c_k$ is its unique maximum in $\ell_k=(a_k,b_k)$. So $c_n>0$.
Obviously
$$
G_n(c_n) \geq G(c_n)
$$ 
by the principle of the extension of the domain.
But clearly $G_n(x)$ is a decreasing function on $(-\infty,b_n]$.
Therefore,
$$
G_n(0) \geq G_n(c_n)
$$
Two last inequalities together show that series $\sum_n G(c_n)$ converges (Widom's condition) if series
$\sum_n G_n(0)$ converges. So we are left to see why the fact that \eqref{12} holds for all $n$ implies that  series
$\sum_n G_n(0)$ converges. 
We write the Poisson integral for $x\in \Omega_n$:
$$
G_{n-1}(x) -G_n(x) = \int_{[b_n, a_{n-1}]} G_{n-1}(y) \,d\omega_{n}(y,x) \,.
$$
Now we use Harnack's inequality to conclude
$$
\tau\,G_{n-1}(0) \leq G_{n-1}(y)\,\,\forall y\in [b_n, a_{n-1}]\,.
$$
Here $\tau$ is an absolute constant. In fact, all points of the segment $[b_n, a_{n-1}]$ are much closer to $0$ than to $\partial \Omega_{n-1}= [b_{n-1}, a_{n-2}]\cup [b_{n-2}, a_{n-3}]\cup...\cup [b_1,a_0]$.
Therefore,
$$
G_{n-1}(0) -G_n(0) \geq \tau \int_{[b_n, a_{n-1}] }G_{n-1}(0) \,d\omega_{n}(y,x) =
\tau\omega_{n}([b_n, a_{n-1}],0)G_{n-1}(0) \geq \frac{\tau}2 G_{n-1}(0) \,.
$$
We get
\begin{equation}
\label{geom}
G_n(0) \leq (1-\frac{\tau}2) G_{n-1}(0)\leq ....\leq C\, (1-\frac{\tau}2)^{n-1}\,.
\end{equation}
We proved the convergence of $\sum G(c_n) \leq \sum G_n(0)$: the Widom property.

Now we prove Lemma \ref{harmonic}.

\begin{proof}
Without loss of generality we can think that $E\subset [1,\infty)$, $0<b'< a<1$. We also may consider special $\Omega= \C\setminus ([b',a]\cup [1,\infty))$, the general case follows immediately by the principle of extension of the domain. Now let $k=b'/a$. Rescale the domain to get $\Omega_k = \C\setminus ([k,1]\cup [1/a,\infty))$. Keep $k$ fixed and let $a\rightarrow 0$. Then obviously $\omega_{\Omega_k}( [1/a,\infty),0)\rightarrow 0$. So for some $a(k)$ 
$\omega_{\Omega_k}( [k,1],0)\geq\frac12$.  Rescaling back, we see that we can choose $k$ as close to $1$ as we wish, and choose $a(k)$ in such a way that
$$
\omega_{\Omega} ([b',a],0)\geq\frac12\,.
$$
We are done.
\end{proof} 

\section{Harmonic measures are weakly reflectionless.} 
 \label{harmo}
 
 In \cite{MPV} the following result is proved:
 Let $E$ be  a compact subset of the real line. Let harmonic measure $\omega$ of $\C\setminus E$ be absolutely continuous with respect to $dx$ on $E$. Then $\omega$ is such that $C_{\omega}(x+i0)$ is $0$ $dx$ (and hence $d\omega$) almost everywhere on $E$.
 
 We show here by a simple way that a more general result holds:
 \begin{theorem}
 \label{anyh}
 Let $E$ be  a compact subset of the real line. Then $\omega$ is such that $C_{\omega}(x+i0)$ is $0$ $dx$ almost everywhere on $E$.
 \end{theorem}
 
 \begin{proof}
 Let $\ell_n$ be complementary intervals to $E$ with $\ell_0$ being the interval containing  infinity. Consider $\Omega_n:= (\C \setminus\R) \cup \ell_1\cup...\cup \ell_n$. Let $b_0$ be the leftest point of $E$, $a_0$ being the rightest point of $E$. We use the notations from above. Let
 $$
 R_n(z) =\int \frac{d\omega_n(t)}{t-z}\,,
 $$
 we know that 
 
\begin{equation}
\label{positive11}
(x-b_0)R_n(x) <0,\,\,\, \text{for}\,\,\, x< b_0, \,\,\,\text{or} \,\,\, x>a_0\,.
\end{equation}
And $(z-b_0) R_n(z)$ is a Nevanlinna function (positive imaginary part in $\C_+$). 
Therefore,
we can define  logarithm of $-(z-b_0) R_n(z)$in the upper half plane, and the  imaginary part of this logarithm has compact support on the real line. Call it $-g_n(x)$. Then (for $z\in\C_+$)
$$
R_n(z) = - \frac1{z-b_0} e^{-\int\frac{g_n(t)dt}{t-z}}\,.
$$
Similarly,
\begin{equation}
\label{R}
R(z) = - \frac1{z-b_0} e^{-\int\frac{g(t)dt}{t-z}}\,.
\end{equation}

Of course compactly supported (on $[b_0,a_0]$) functions $g_n$ have values in $[0,\pi]$, and $g_n\rightarrow g$ weakly in $L^2$. In fact, $R_n(z)$ converges to $R(z)$ uniformly on compact sets in $\C\setminus\R$ (this is just weak convergence of $\omega_n$ to $\omega$). Then $\int\frac{g_n(t)dt}{t-z}\rightarrow \int\frac{g(t)dt}{t-z}$ uniformly on compact sets. At the same time $g_n$ and $g$ are compactly supported and have $L^{\infty}$ norm bounded by $\pi$. Therefore, indeed $g_n\rightarrow g$ weakly in $L^2$. But then certain convex combinations of $g_n$ converge to $g$ a.e. with respect to $dx$. All these convex combinayions are identically equal to $\frac{\pi}2$ on $E$. We conclude that $g=\pi/2$ a.e. on E with respect to Lebesgue measure. Now formula \eqref{R} shows that $C_{\omega}(x+i0) =0$ Lebesgue a.e. on E. We are done.
 
 \end{proof}
 
 \section{Singular continuous components of weakly reflectionless measures.}
 \label{cont}
 
 Now we are  going to construct the closed set $E= [b_0,a_0]\setminus \cup (a_j, b_j)$ such that function
$$
R(z) = -\frac{1}{\sqrt{(z-a_0)(z-b_0)}}\Pi_{j=1}^{\infty}\frac{z-x_j}{\sqrt{(z-a_j)(z-b_j)}}=:\int_E \frac{d\mu(t)}{t-z}
$$
has a singular continuous component in measure $\mu$.
We construct $E$ inductively. Let $b_0=-1, a_0=1$. Let us choose very fast decreasing to zero sequence $\e_n$, and let sequence $L_n$ be such that $L_{n+1}/L_n < \e_n$. The set $E$ will be comprised of closed segments and the limit points. 

\noindent Step 1.

First two segments
$s_1:=[b_0,c_0], s_2:=[d_0,a_0]$ are such that for $z\in n_1:= [-L_1/2,L_1/2]$ we have
$$
1-\e_1<\Big|\frac{z-d_0}{z-a_0}\Big|^{\frac12} \Big|\frac{z-c_0}{z-b_0}\Big|^{\frac12}<1\,.
$$ 
Given a segment $s=[\alpha,\beta]$ we put $r_s(z):=\Big|\frac{z-\alpha}{z-\beta}\Big|^{\frac12}$.
We put $p_1=c_0, p_2= d_0$.  We choose now two symmetric segments $s_3, s_4$ such that $s_3<0, s_4>0$ and
\begin{itemize}
\item $r_{s_i}(0) \in (1-\e_2, 1+\e_2),\. i=3,4$;
\item Green's function $G_4$ of $\Omega_4:=\C\setminus \cup_{i=1}{4}s_i$ is smaller than $\e_2$ between $s_3, s_4$.
\end{itemize}
We know by Section \ref{example} that we just need to have segments $s_3, s_4$ close enough to zero.

Now we choose interval $n_2=(ln_2,rn_2)$ centered at zero of length $\leq L_2$ and such that
$$
r_{s_i}(z) \in (1-\e_2, 1+\e_2),\. i=3,4,\,\forall z\in n_2\,.
$$

We put $p_3$ at the right endpoint of $s_3$ and $p_4$ at the left endpoint of $s_4$.

To finish step 1 we choose segment $s_5$ centered at zero, inside $n_2$ and of such a small length $2\ell_1$ that
$$
r_{s_5}(z) \in (1-\e_3, 1+\e_3)\,\, z=rn_2/2, ln_2/2\,.
$$

Put $E_5:=\cup_{i=1}^{5}s_i$.
Notice that 
\begin{itemize}
\item Green's function $G_5$ of $\Omega_5:=\C\setminus E_5$ is smaller than $\e_2$ between $s_3, s_4$.
\item Measure $\mu_5$ built by our formula on $E_5$ by assignment of four points $p_1,...,p_4$ has very large mass on $s_5$, in fact
\end{itemize}
\begin{equation}
\label{s5large}
\mu_5(s_5) \geq (1-\e_1)(1-\e_2)^2\int_{-\ell_1}^{\ell_1}\frac{dx}{\sqrt{\ell_1^2-x^2} }\,.
\end{equation}

\noindent Step 2.

Now we choose interval $n_3=(ln_3,rn_3), n_4=(ln_4,rn_4)$ centered at $ln_2/2$ and $rn_2/2$ of length $\leq L_3$ and such that
\begin{equation}
\label{s5vnen2over4}
r_{s_5}(z) \in (1-\e_3, 1+\e_3),\, \,\forall z\in n_3\cup n_4\cup (\R\setminus (ln_2/4, rn_2/4))\,.
\end{equation}
This is only one requirement on $n_3, n_4$. We will list another.
Put $p_5,p_6$ as left and right endpoints of $s_5$.
Choose pair of segments $s_6, s_7$ symmetrically around $ln_2$, and $s_8, s_9$ symmetrically around $rn_2$
We do this so that
\begin{itemize}
\item $r_{s_i}(ln_2) \in (1-\e_3, 1+\e_3),\. i=6,7$;
\item $r_{s_i}(rn_2) \in (1-\e_3, 1+\e_3),\. i=8,9$;
\item $s_6, s_7, s_8, s_9 \subset n_2$;
\item $r_{s_i}(z) \in (1-\e_3, 1+\e_3),\. i=6,7,8,9,\,\,for all z\in \R\setminus n_2$;
\item Green's function $G_9$ of $\Omega_9:=\C\setminus \cup_{i=1}{9}s_i$ is smaller than $\e_3$ between $s_6, s_7$ and between $s_8,s_9$.
\end{itemize}
We know by Section \ref{example} that we can reconcile the first four items with the fifth one: we just need to have segments $s_6, s_7$ with small $r_{s_i}(ln_2)$ , $i=6,7$, close enough to $ln_2$ and $s_8, s_9$ with small $r_{s_i}(rn_2)$ , $i=8,9$,  close enough to $rn_2$ .

Now we have the second requirement on lengths of $n_3, n_4$:
\begin{itemize}
\item $r_{s_i}(z) \in (1-\e_3, 1+\e_3),\. i=6,7,\,\,\forall z \in n_3\cup (\R\setminus n_2)$;
\item $r_{s_i}(z) \in (1-\e_3, 1+\e_3),\. i=8,9,\,\,\forall z \in n_4\cup (\R\setminus n_2)$;
\end{itemize}

To finish step 2  we denote by $ln_{51}, rn_{51}$ points which a mid-points of the left and the right halves of segment $n_3$ correspondingly. By $ln_{52}, rn_{52}$ points which a mid-points of the left and the right halves of segment $n_4$ correspondingly. 

Given interval $J$  and number $\lambda>0$, we denote (as usual) by $\lambda J$ the interval with the same center and the lenghth $\lambda |J|$.

We choose segments $s_{51}, s_{52}$ centered at $ln_2/2$, $rn_2/2$ correspondingly, inside $n_3$, $n_4$ correspondingly, and of such a small length $2\ell_2l$ that
\begin{equation}
\label{s51vnen4over4}
r_{s_{51}}(z) \in (1-\e_4, 1+\e_4)\,\, x\in \R\setminus \frac12 (ln_{51}, rn_{51}),\,\,\text{in particular for}\,\,z=ln_{51}, rn_{51};
\end{equation}
\begin{equation}
\label{s52vnen4over4}
r_{s_{52}}(z) \in (1-\e_4, 1+\e_4)\,\,  x\in \R\setminus \frac12 (ln_{52}, rn_{52}),\,\,\text{in particular for}\,\,z=ln_{52}, rn_{52}\,.
\end{equation}

Put $E_{11}:=\cup_{i=1}^{9}s_i\cup s_{51}\cup s_{52}$.

Now we have $10$ complementary intervals to $E_{11}$ on $[-1,1]$. Assign $p_5, p_6$ to left and right endpoints of segment $s_5$. This will make measure $\mu_{11}$ very small on $s_5$. Recall that $\mu_5$ was quite large on $s_5$---so what we are doing is the process of removal of mass from $s_5$. Where this mass will disappear? To understand this first let us assign the points $p_7,...,p_{10}$. 

We have a complementary interval between segments $s_6$ and $s_{51}$. Put $p_7$ in its left endpoint (=right endpoint of $s_6$). 

We have a complementary interval between segments $s_{51}$ and $s_{7}$. Put $p_8$ in its right endpoint (=left endpoint of $s_7$). 

We have a complementary interval between segments $s_8$ and $s_{52}$. Put $p_9$ in its left endpoint (=right endpoint of $s_8$). 

We have a complementary interval between segments $s_{52}$ and $s_{9}$. Put $p_{10}$ in its right endpoint (=left endpoint of $s_9$). 

This assignment fixes function $R_{11}(z)=\int_{E_{11}}\frac{d\mu_{11}(x)}{x-z}$.

As  a result of the construction mass almost disappears from $s_5$:
\begin{equation}
\label{s5small}
\mu_{11}(s_5) \leq (1+\e_1)(1+\e_2)^2(1+\e_3)^4\int_{-\ell_1}^{\ell_1}\sqrt{\ell_1^2-x^2}\,dx \asymp \ell_1^3\,.
\end{equation}

Let us see that it reappears on $s_{51}, s_{52}$ in almost equal parts. Let us denote temporarily the endpoints of $s_{51}$ by $a,b$, endpints of $s_5$ by $c,d$ and endpoints $s_{52}$ by $e,f$. We do not need endpoints of $s_k, k=1, 2,...,9$ as $r_{s_k}(z) \in (1-\e_i, 1+e_i)$ for $i=1,2,3$ for $z\in s_{51}$ by construction.
Then for $x\in s_{51}$ we clearly have
\begin{equation}
\label{schet}
\frac{d\mu_{11}(x)}{dx} /[ \frac{\sqrt{(c-x)(d-x)}}{\sqrt{(e-x)(f-x)}}\frac{1}{\sqrt{(b-x)(x-a)}}] \in ((1-\e_1)(1-\e_2)^2(1-\e_3)^2, (1+\e_1)(1+\e_2)^2(1+\e_3)^2)
\end{equation}
But from \eqref{s5vnen2over4} and \eqref{s52vnen4over4} we get

\begin{equation}
\label{schet1}
\frac{\sqrt{(c-x)(d-x)}}{\sqrt{(e-x)(f-x)}}\in \frac12 ( (1-\e_3)^2(1-\e_4)^2, (1+\e_3)^2(1+\e_4)^2)\,.
\end{equation}
 Combining \eqref{schet}, \eqref{schet1} we get
 
 \begin{equation}
 \label{mu11ons51}
 \mu_{11}(s_51) \geq \frac12  \Pi_{i=1}^4(1-\e_i)^{2^i} \int_{-\ell_2}^{\ell^2}\frac{dx}{\sqrt{\ell_2^2-x^2}} \asymp \frac12 \mu_5(s_5)\,.
 \end{equation}
 Symmetrically
 
 \begin{equation}
 \label{mu11ons52}
 \mu_{11}(s_52) \geq \frac12  \Pi_{i=1}^4(1-\e_i)^{2^i} \int_{-\ell_2}^{\ell^2}\frac{dx}{\sqrt{\ell_2^2-x^2}} \asymp \frac12 \mu_5(s_5)\,.
 \end{equation}

Notice that 
\begin{itemize}
\item Green's function $G_{11}$ of $\Omega_{11}:=\C\setminus E_{11}$ is smaller than $\e_3$ between $s_6, s_7$ and between $s_8,s_9$.
\item Measure $\mu_{11}$ built by our formula on $E_{11}$ by assignment of ten points $p_1,...,p_{10}$ has very large mass on $s_{51}, s_{52}$, in fact these masses are
$ (\frac12 -\frac18) \mu_5(s_5) \,.$
\end{itemize}

\noindent Renaming. We use name $n_{51}$ for $n_3$, name $n_{52}$ for $n_4$ and even $n_5$ for $n_2$.  We call $s_6$---$Ls_{51}$, $s_7$---$Rs_{51}$, $s_8$---$Ls_{52}$, $s_9$---$Rs_{52}$. We rename $E_5$ into $E^1$, it has $5$ segments, $\Omega^1 := \S \setminus E^1$ (it was called $\Omega_5$ before).  Also $E^2$ is now is $E_{11}$ it has $5+ 2 + 2^2$ segments, $\Omega_2=\C\setminus E^2$.

\noindent Next steps. We are going to repeat previous steps. We put a very small $n_{511}$ centered at $ln_{51}$, $n_{512}$ centered at $rn_{51}$,
$n_{521}$ centered at $ln_{52}$, $n_{522}$ centered at $rn_{52}$.  (These $ln_{5i}. rn_{5j}, i,j=1,2$ are midpoints of left and right halves of $n_{5i}$.
We build $Ls_{5ij}, Rs_{5ij}, i, j= 1,2$ as before. And then we build $s_{5ij}, i,j=1,2$.
We get $E^3$ with $5+ 2 + 2^ 2 + 2^2 + 2^3$ segments,
then $E^4$ with $5+ 2 + 2^2 + 2^2 + 2^3 + 2^3+ 2^4$  segments.
Measures are called $\mu^k$ now, so $\mu^1$ is $\mu_{5}$ and $\mu^2$ is $\mu_{11}$. Et cetera....
We already saw that
\begin{equation}
\label{half2}
\mu_{2}(s_{5i} )\geq (\frac12-\frac18)\mu^1(s_5),\,\,\forall i=1,2\,.
\end{equation}

By construction will also have

\begin{equation}
\label{half3}
\mu_{3}(s_{5ij} )\geq (\frac12-\frac18)(\frac12-\frac1{16})\mu^1(s_5),\,\,\forall i,j=1,2\,.
\end{equation}

Similarly,
\begin{equation}
\label{half4}
\mu_{4}(s_{5ijk} )\geq (\frac12-\frac18)(\frac12-\frac1{16})(\frac12-\frac1{32})\mu^1(s_5),\,\,\forall i,j=1,2\,.
\end{equation}

Et cetera...

So on the union of $n$-th generation $s_{5i_1...i_n}$ we still have measure
$\mu^{n+1}$ having a mass bigger than numerical positive constant $c_0$.

On the other hand $\mu^{n+2}, \mu^{n+3},....$ are small on the union of these intervals.

Therefore the limit measure $\mu$ has mass $c_0$ {\bf not} on the union of constructed countably many intervals of all generations. So this measure is of the {\bf limit set} of these intervals, so this measure has a singular component of mass at least $c_0$.

Why this singular component does not have point component?

This is because we always divide the measure of each segment $s_{5i_1...i_n}$ into two almost equal parts. Hence one can easily see that

\begin{equation}
\label{nopints}
\lim_{|I|\rightarrow 0} \mu(I) =0\,.
\end{equation}

Of course we need
$$
\sum 2^i\e_i <\infty\,.
$$

\noindent {\bf Remark 1.} We can also construct Widom domain $\Omega=\C\setminus E$ and distribute $p_j$ in such a way that measure $\nu$ in the representation
$$
-1/R(z) -z-c=\int_E\frac{d\nu(x)}{x-z}
$$
has a continuous singular part, and it is still such that $R(x+i0)$ is purely imaginary $dx$ a.e. on $E$.

Next is a curious remark about {\bf homogeneous} sets $E$! Recall that for  homogeneous sets $E$ reflectionless measures $\mu$ always give rise to $H^1$ functions $R$ and all reflectionless measures supported on homogeneous set $E$ and having it as its abs.  continuous spectrum cannot have any singular component! However,

\noindent {\bf Remark 2.} We can also construct  domain $\Omega=\C\setminus E$ with {\bf homogeneous} $E$ and distribute $p_j<q_j$ in such a way that after constructing $R$ by $p_j$ and $Q$ by $q_j$ we get $R/Q$ to be a Nevanlinna function and the measure $\nu$ in the representation
$$
\frac{R(z)}{Q(z)}=\int_E\frac{d\nu(x)}{x-z}
$$
has a continuous singular part.


\begin{thebibliography}{99}

\bibitem[MPV]{MPV} {\sc M. Melnikov, A. Pltoratskii, A. Volberg}, {\em Uniqueness theorems for Cauchy integral}, arXiv, 2007.

\bibitem[SoYu]{SoYu} {\sc M. Sodin, P. Yuditskii},  {\em Almost periodic Jacobi matrices with homogeneous spectrum, infinite dimensional Jacobi inversion, and Hardy spaces of character automorphic functions.} J. of Geom. Analysis, {\bf 7} (1997), 387--435.

\bibitem[Z]{Z} {\sc M. Zinsmeister}, {\em Espaces de Hardy et domaines de Denjoy}, Ark. Mat. {\bf 27} (1989), 363--378.


\end{thebibliography}
\end{document}